\documentclass[journal]{IEEEtran}

\ifCLASSINFOpdf
  \usepackage[pdftex]{graphicx}
  \DeclareGraphicsExtensions{.pdf,.jpeg,.png}
\else
  \usepackage[dvips]{graphicx}
  \DeclareGraphicsExtensions{.eps}
\fi

\usepackage{siunitx}

\usepackage{graphicx,rotating,booktabs}
\usepackage[verbose]{placeins}
\usepackage[english]{babel}
\usepackage{blindtext}
\usepackage[most]{tcolorbox}
\usepackage{epstopdf}
\usepackage[export]{adjustbox}

\usepackage{color,soul}
\newcommand{\argminD}{\arg\!\min} 
\usepackage{amssymb}
\usepackage{mathtools}
\usepackage{algorithm}
\usepackage{algorithmic}
\usepackage{nomencl}

\begin{document}

\title{Printed Texts Tracking and Following for a Finger-Wearable Electro-Braille System Through Opto-electrotactile Feedback }

\author{Mehdi~Rahimi,~Yantao~Shen$^{*}$,~Zhiming~Liu,~and Fang~Jiang
\thanks{M. Rahimi, PhD in Biomedical Engineering, is with the Department of Chemistry, University of Colorado-Denver, Denver, CO, 80210 USA.  E-mail: mehdi.rahimi@ucdenver.edu
}
\thanks{Y. Shen, PhD and Z. Liu, PhD are with the Department
of Electrical and Biomedical Engineering, University of Nevada-Reno, Reno,
NV, 89557 USA.  E-mail: ytshen@unr.edu. 
}
\thanks{F. Jiang, PhD is with the Department
of Psychology, University of Nevada, Reno, Reno,
NV, 89557 USA e-mail: fangj@unr.edu.}
\thanks{
This research work was partially supported by NSF CAREER Award CBET\#1352006 and the National Eye Institute of the National Institute of Health under award R01EY026275.}
}
\maketitle

\begin{abstract}
This paper presents our recent development on a portable and refreshable text reading and sensory substitution system for the blind or visually impaired (BVI), called Finger-eye. The system mainly consists of an opto-text processing unit and a compact electro-tactile based display that can deliver text-related electrical signals to the fingertip skin through a wearable and Braille-dot patterned electrode array and thus delivers the electro-stimulation based Braille touch sensations to the fingertip. To achieve the goal of aiding BVI to read any text not written in Braille through this portable system, in this work, a Rapid Optical Character Recognition (R-OCR) method is firstly developed for real-time processing text information based on a Fisheye imaging device mounted at the finger-wearable electro-tactile display. This allows real-time translation of printed text to electro-Braille along with natural movement of user's fingertip as if reading any Braille display or book. More importantly, an electro-tactile neuro-stimulation feedback mechanism is proposed and incorporated with the R-OCR method, which facilitates a new opto-electrotactile feedback based text line tracking control approach that enables text line following by user fingertip during reading. Multiple experiments were designed and conducted to test the ability of blindfolded participants to read through and follow the text line based on the opto-electrotactile-feedback method. The experiments show that as the result of the opto-electrotactile-feedback, the users were able to maintain their fingertip within a $2mm$ distance of the text while scanning a text line. This research is a significant step to aid the BVI users with a portable means to translate and follow to read any printed text to Braille, whether in the digital realm or physically, on any surface.
\end{abstract}

\IEEEpeerreviewmaketitle

\section{Introduction and Related Work}
\IEEEPARstart{T}{}he number of people of all ages globally that are visually impaired is estimated to be 285 million and is increasing \cite{pascolini2012global}. This is mainly due to increase in screen time usage and also the global trend of aging. To help their reading, the two popular options for this population are either the Braille language or audio means. There are challenges with using the audio as 90\% of teachers of blind students agree that technology (such as text-to-audio) should be used only as a \textit{supplement} to Braille rather than as a \textit{replacement}  \cite{wittenstein1996teachers}. Also, based on another study, only 10\% of blind people who did not know Braille could find a job \cite{national2009braille}. These facts show the importance of the Braille in comparison to audio solutions.

The primary benefit of Braille is that it allows users to read in their preferred manner with tactile feeling such as
skimming the text, searching for bold or highlighted text and etc. But it also introduces new limitations such as the availability of the Braille books, their size, weight, and cost. As an example, the Webster's Collegiate Dictionary Braille copy is over 75 volumes, each volume over three inches thick and costs over five thousand dollars \cite{hedgpeth2006icare}. 

Conventional refreshable braille displays are electro-mechanical approaches to translate text to Braille and can solve the problems associated with the traditional Braille systems \cite{loewen2002fostering}. However, there are still challenges involved in using a heavy or bulky mechanical device. 

Numerous research has been done in the field of text-to-Braille research including devices such as \textit{Optacon} \cite{kendrick2005optacon}, \textit{OrCam} \cite{na2012user}, \textit{FingerReader} \cite{shilkrot2015fingerreader}, 
\textit{BrailleRing} \cite{zagler2018braillering}, and \textit{HandSight} \cite{froehlich2015handsight} where each has proposed different solutions to the problem of converting the written text to Braille in real-time. The use of a camera connected to an Optical Character Recognition (OCR) program to extract the text information of the picture is a common practice among all of these approaches. In these devices, the output is given to the user by either haptic feedback or audio. 

Our research has introduced a new approach by developing an electro-tactile display that can be used as a refreshable Braille display without any mechanical parts.
An electro-tactile stimulation based display has numerous advantages over the the mechanical approaches. These include being significantly more compact, being easier to fabricate, and also, having lower power consumption. The absence of any mechanical parts will lead to ease of repair and lower production costs as well.

In addition, a comparison between mechanical stimulation of the skin and electro-tactile stimulation has demonstrated various benefits for the electrical approaches \cite{bach2003seeing}. One significant factor would be the undeniable shorter time that an electrical signal needs for any changes, whereas a mechanical approach is more limited by the constraints of the moving parts of the motors and other segments of the system \cite{antfolk2013sensory}. The skin reaction to electrical stimulation is also considerably faster and more accurate \cite{szeto1982electrocutaneous}\cite{kaczmarek1991electrotactile}.  Electrical stimulation generally has a higher convenience level, better efficiency, and more flexibility for the purpose of giving sensory information along with less noise and lighter weight of the system \cite{kaczmarek1995sensory}.

In this research, our electro-tactile display is built using an electrode array of contacts to deliver an electrical signal to the skin \cite{liu2016finger}. The main concepts and mechanism of an electro-tactile display have been investigated mostly by Kaczmarek and Tyler \cite{bach2003seeing}\cite{kaczmarek2017portable}. Their work was mainly focused on a display to be placed on the tongue resulting in developing \textit{BrainPort} head-mounted system \cite{danilov2005brainport} and electro-tactile displays for the fingertip \cite{kaczmarek1994electrotactile}. Other electro-tactile locations on the body have been also examined. Kajimoto, proposed an electro-tactile display for the forehead \cite{kajimoto2006forehead} and palm of the hand \cite{kajimoto2016electro}\cite{kajimoto2014hamsatouch}. He also explored the possibility of using a display for the fingertip \cite{kajimoto2004electro}\cite{kajimoto2002electrocutaneous}. One of the benefits of using the fingertip as the location for the display is that the skin has the highest spatial resolution in human body \cite{kajimoto2006forehead}. Among electro-tactile studies mentioned above, most of them placed the return electrodes in a very close proximity of the stimulating electrodes \cite{kajimoto2003psychophysical} although the two-point discrimination threshold (TPDT) of the skin may limit the remaining area on the skin \cite{solomonow1977electrotactile}. Additional concern in such an arrangement may cause the electrical current leak to the surrounding electrodes \cite{sato2007electrotactile}. In addition, supplying the user with an appropriate stimulation based on the properties of the skin by electro-tacile displays is a challenging topic \cite{dorgan1999model}\cite{gregory2009towards}. This has been shown to be very complicated and problematic as the model is time-varying in each part of the skin \cite{rahimi2019non}. Some researchers have tried to mitigate it by controlling the pulse width of the signal \cite{kajimoto2012electrotactile}, while others tried using a mapping approach as the solution \cite{tyler2009spatial}. A normalized mean threshold map is used in the research \cite{moritz2017perceived}\cite{wilson2012lingual}. 

For our electro-tactile display, a new adaptive spatial mapping strategy was developed to supply the appropriate stimulation feeling \cite{PaperMAP2018NOTPUBLISHED}.
In this approach, we chose to put the active electrode array slightly distal to the fingertip vortex. A bigger single return electrode is placed on the palm of the hand as the return electrode. This configuration can offer the advantages of allowing more active electrodes to fingertip so as to best match the density of the tactile receptors in situ \cite{shimoga1993survey}\cite{johansson2009coding}. Having the return electrode in another part of the hand, other than the fingertip, will lead to better assessing of the feeling of sensations, whereas placing the return electrode in the close proximity can interfere in this process. The characteristics of the signal, such as the voltage, frequency and duty cycle have been also  studied before to select the proper sensation for the user. A comfortable signal was chosen as having a frequency of $30Hz$ and duty cycle of $10\%$, which agrees with the voltage setting at the detection threshold level \cite{PaperET2018NOTPUBLISHED}. In this research, we follow those achievements to pursue the most comfortable signal/stimulation for the user.

Based on our electro-tactile display array and methods, this work focuses on the feedback that the user needs to perceive so as to maintain their fingertip on the text line while scanning the printed texts. The feedback mechanism measures the appropriate distance to the line and gives the corresponding stimulation feedback to the user as to adjust the fingertip up or down on the text line. These feedbacks are simultaneously given through our electro-tactile display array on user's fingertip. A preliminary investigation of users recognition of the directional commands was done in our previous research \cite{PaperMov2018NOTPUBLISHED}. It was found that the users had more difficulty in identifying the moving pattern if the moving direction is horizontal rather than vertical. This was further verified by combining the results of the movements by disregarding the effects of the diagonal patterns \cite{PaperMovMap2019NOTPUBLISHED}. Based on these findings, the directional feedback in this work are more related to vertical movements. Our approach is verified that it helps keep the users' fingertip within $2mm$ distance of the followed text line, which ensures providing an exceptional accuracy in tracking the text line for the BVI users. 

The contribution and novelty of this work lie in 1) developed a portable and refreshable text reading and sensory substitution system for the blind or visually impaired (BVI). The system mainly consists of an opto-text processing unit and a compact electro-tactile based display that can deliver text-related electrical signals to the fingertip skin through a wearable and Braille-dot patterned electrode array and thus delivers the electro-stimulation based Braille touch sensations to the fingertip; 2) developed a Rapid Optical Character Recognition (R-OCR) method for real-time processing text information based on a Fisheye imaging device mounted at the finger-wearable electro-tactile display. This allows real-time translation of printed text to electro-Braille along with natural movement of user's fingertip as if reading any Braille Display or book; 3) proposed an electro-tactile neuro-stimulation feedback mechanism and incorporated it with the R-OCR method to facilitate a new opto-electrotactile feedback based text line tracking control approach that enables text line following by user fingertip during reading; 4) Multiple experiments were designed and conducted to validate the performance of the methods and systems as well as testing the ability of blindfolded participants in using the system. This work is the first-ever synergetic effort to tackle the challenges from three angles (i.e., perception, computer vision, and human-device interaction) that will produce a holistic and intelligent solution for a complete reading aid for BVI.

This paper is structured into the following sections: in Section \ref{sec:method}, we introduce the custom-built electro-tactile based Braille system and methods that are used in this study; Section \ref{sec:validation} describes the experimental procedures and set-up for validating the developed system and methods; Section \ref{sec:discussion} presents the results obtained from the analysis and includes related discussion about them; in Section \ref{sec:conclusion} we conclude the work.

\section{The System and Methods}
\label{sec:method}

\subsection{Hardware and Systems}

This research is capable of significantly aiding the BVI with a wearable and mobile device to translate any text to Braille and/or speech, whether in the digital realm or physically, on any surface. 

The interactions of the users with the system are through a Braille display. The developed Braille display is based on an 8-dot Braille system similar to the Gardner-Salinas Braille codes commonly known as GS8 \cite{kacorri2013design}. This 8-dot system has two extra dots on the bottom row. This gives an added benefit of being able to display special characters in mathematics and science such as Greek letters, mathematical symbols and also Unicode characters. All these added benefits are alongside supporting the traditional 6-dot Braille. 

\begin{figure}[htbp]
  \centering
  \includegraphics[width=0.65\columnwidth,keepaspectratio, trim=0mm 0mm 0mm 0mm, clip=true]{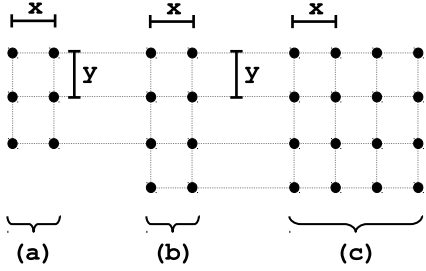}
  \caption[]{(a) the traditional 6-dot Braille; (b) the 8-dot Braille; (c) our proposed 16-dot Braille display. The length of $x$ is $2.29mm$ and $y$ is $2.54mm$.}
  \label{fig:6_8_16_dot_braille}
\end{figure}

We extended this 8-dot system to a 16-dot display by adding two more columns on each side. We use these extra 8 dots for sending commands required for controlling the movements of the fingertip on the text line. These different setups are shown in Fig. \ref{fig:6_8_16_dot_braille}.

Instead of using a PC for the image processing and controlling applications, a single board computer (SBC) was used for the purpose of portability of the system. The SBC used in this research was chosen as a Raspberry Pi 3 (RPi3) with 1GB RAM and a CPU frequency of 1.2GHz. An upgrade to RPi3+ was not considered because of the insignificant improvement, but an UDOO board may be used in the future to improve the processing power. The power of the RPi3 was determined sufficient at this stage of the research.

The camera used in this research is an I2C camera with CSI conversion. Fish-eye cameras are special cameras that have a much wider field of view than the conventional cameras. Therefore, they have been used increasingly in many applications such as computer vision, robotics, surveillance, etc. But the large field of view pays a price for non-linear distortions introduced around the boundary of the image captured by fish-eye cameras. 
\\
The fish-eye camera matrix is the same as conventional camera, which can be modeled as shown in (\ref{eq:fisheye}).
\begin{equation}
\label{eq:fisheye}
\begin{bmatrix}
    f_x       & 	0 		& c_x  \\
    0		  & 	f_y     & c_y  \\
    0         & 	0 		& 1  \\
\end{bmatrix}
\end{equation}
but with different distortion coefficients. Let $(X, Y, Z)$ be a point in the world reference coordinates. The image coordinates after projection is $(u, v)$ without considering distortion. This is shown in (\ref{eq:fisheye2}).
\begin{equation}
\label{eq:fisheye2}
\begin{gathered}
\begin{bmatrix}
    x  \\
    y  \\
    z  \\
\end{bmatrix}
=R
\begin{bmatrix}
    X  \\
    Y  \\
    Z  \\
\end{bmatrix}
+T\\
x^{'}=x/z\\
y^{'}=y/z\\
u=f_x * x^{'} + c_x\\
v=f_y * y^{'} + c_y
\end{gathered}
\end{equation}
where $R$ and $T$ are the extrinsic parameters: rotation and translation. When the distortion is considered for fisheye camera, $(u, v)$ can be obtained following the rationale in (\ref{eq:dis}).
\begin{equation}
\label{eq:dis}
\begin{gathered}
r^2=x^{'2}+y^{'2}\\
\theta=tan^{-1}(r)\\
\theta^{'}=\theta(1+k_1 \theta^{2} + k_2 \theta^{4} + k_3 \theta^{6} + k_4 \theta^{8})\\
x^{'}=(\theta^{'}/r) * x \\
y^{'}=(\theta^{'}/r) * y \\
u=f_x * x^{'} + c_x\\
v=f_y * y^{'} + c_y
\end{gathered}
\end{equation}

This results in an un-distortion image after calibration. It can be noticed  that the image area located far from the center of image is also been un-distorted perfectly.

\begin{figure}[htbp]
  \centering
  \tcbox[top=0pt,left=0pt,right=0pt,bottom=0pt]
  {\includegraphics[width=0.85\columnwidth,keepaspectratio, trim=13mm 10mm 13mm 10mm, clip=true]{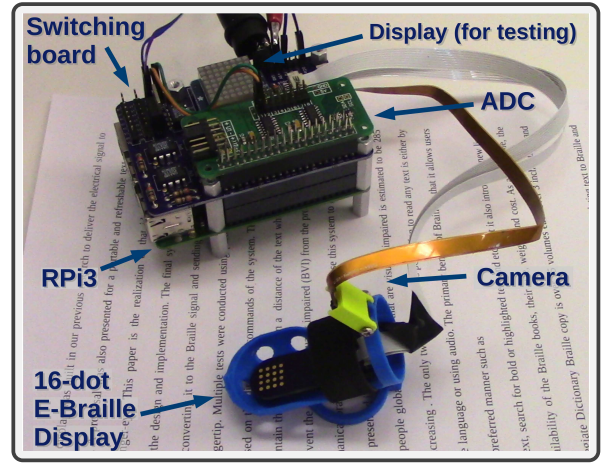}}
  \caption[]{The system and its components are shown here. The RPi3 can work as a stand-alone system without any dependency on a PC. The ADC and the switching board are to control the voltage and current passing through the skin.}
  \label{fig:system}
\end{figure}

The system and its components are shown in Fig. \ref{fig:system}. The developed E-braille display can also be seen in this figure.

The system supplies the fingertip with a voltage necessary to produce almost $30\mu A$. This voltage ranges between $60V$ to $100V$ for different users and different conditions of the skin extensively explained in our previous research.

\subsection{Rapid OCR for Fish-eye Camera}

\begin{figure}[htbp]
  \centering
  \tcbox[top=0pt,left=0pt,right=0pt,bottom=0pt]
  {\includegraphics[width=0.8\columnwidth,keepaspectratio]{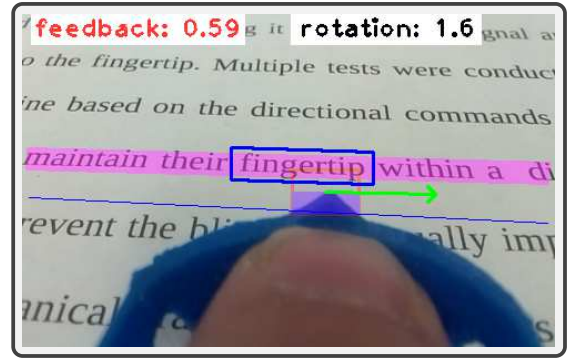}}
  \caption[]{The camera view is shown here. The distance from the line is shown as the "feedback" parameter and the angle is shown with the "rotation" parameter. The image processing algorithm has found the word "fingertip" in this picture and it was shown in a box. This boxed section is later corrected and sent to the OCR.}
  \label{fig:camera_view}
\end{figure}

The camera continuously scans the text for processing. Fig \ref{fig:camera_view} shows the view field of the camera. 

There are multiple preparations steps for the camera.
When camera capture images from non-planar objects, there are usually distortions in the texture image of curved surface. For better performance of image recognition and OCR, a transformation needs to be applied to flatten such images. The common image transformation such as perspective and affine transformation can work correctly only for planar images. Here, We can use thin plate splines (TPS) as the non-rigid transformation model.
The name thin plate spline refers to a physical analogy involving the bending of a thin sheet of metal. Just as the metal has rigidity, the TPS fit resists bending also, implying a penalty involving the smoothness of the fitted surface. Suppose $x$ is two dimensional. The TPS fits a mapping function $f(x)$ between corresponding point-sets ${yi}$ and ${xi}$ that minimizes the following energy function. The relation is shown in (\ref{eq:tps}).
\begin{equation}
\label{eq:tps}
E_{tps}(f)=\Sigma_{i=1}^K \Vert y_i - f(x_i) \Vert ^2
\end{equation}
The smoothing variant uses a tuning parameter $\lambda$ to control how non-rigid is allowed for the deformation, which balances the aforementioned criterion with the measure of goodness of fit. Thus the minimization becomes:
\begin{equation}
\label{eq:min}
\begin{gathered}
E_{tps,smooth}(f) = \Sigma_{i=1}^K \Vert y_i - f(x_i) \Vert ^2 +\\
\lambda \int \int [(\dfrac{\partial^2 f}{\partial x_1^2})^2+ 2 (\dfrac{\partial^2 f}{\partial x_1 \partial x_2})^2 + (\dfrac{\partial^2 f}{\partial x_2^2})^2] dx_1 dx_2
\end{gathered}
\end{equation}
Following (\ref{eq:min}) results in image flattening from cylindrical surface to planar surface.
\\
Another challenge is the blurring or degradation of an image which can be caused by the movement of camera during the image capture process or out-of-focus optics. A blurred image can be approximately described by:
\begin{equation}
\label{eq:blur}
y=h\circledast x + n
\end{equation}
here $y$ is the blurred image, $x$ is the original image, $n$ is the additive noise that is introduced during image acquisition, $h$ is the point spread function (PSF), and $\circledast$ is the convolution operator. The blurring is typically modeled as the convolution of a PSF with a hypothetical sharp input image, where both the sharp input image (which is to be recovered) and the PSF are unknown. In almost all cases, there is insufficient information in the blurred image to uniquely determine a plausible original image. 
\\
A solution for de-blurring considering (\ref{eq:blur}),  is blind deconvolution that allows the recovery of the target scene from a blurred image in the presence of a poorly determined or unknown PSF. The goal of blind deconvolution is to infer both $f$ and $h$ given a single input $g$. The approach here is a maximum-a-posterior (MAP) estimation, seeking a pair $(\hat x, \hat h)$  minimizing:
\begin{equation}
\label{eq:map}
\begin{gathered}
(\hat x, \hat h) = \argminD_{x,h} \lambda \Vert x \circledast h - g \Vert ^2 + \\
\Sigma_{i} \mid g_{x,i}(x) \mid ^ a + \mid g_{y,i}(x) \mid ^ a
\end{gathered}
\end{equation}
Solving (\ref{eq:map}) results is a de-blurred image using the blind deconvolution.  

After having a suitable image, the first step is to check the orientation of the document. The document might have been slightly rotated during acquiring images. It is necessary to detect the skew angle and to rotate the text to be up-right before the subsequent procedures. The popular technique for line detection is the Standard Hough Transform (SHT). 
However, performing SHT is time consuming. Probabilistic Hough Transform (PHT) is an optimization of SHT, which does not consider all the points, instead, it takes only a random subset of points that is sufficient for our line detection subroutine.

\begin{figure}[htbp]
  \centering
  \includegraphics[width=\columnwidth,keepaspectratio, trim=6mm 6mm 6mm 6mm, clip=true]{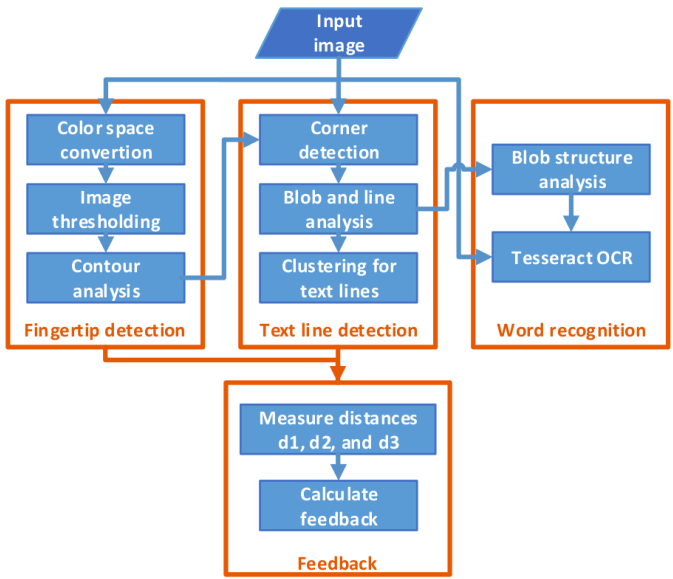}
  \caption[]{The diagram of image processing algorithm is shown here. The algorithm consists of four main parts: fingertip detection, text line detection, word recognition, and feedback for finger movement.}
  \label{fig:img-diagram}
\end{figure}

When the BVI user scans the text, multiple text lines can be captured in the camera's field of view. It is therefore needed to know which word is the center of focus and is going to be recognized. For the BVI users, their fingertip can be assumed as a reference for finding words. Therefore, it is necessary to detect the fingertip in the image and obtain the position of fingertip relative to the text lines. Based on these information, we can design the feedback scheme to guide the BVI users to move their finger on the paper in the right direction. The algorithm of image processing consists of four parts: fingertip detection, text line detection, word recognition, and feedback for finger movement. The algorithm is shown in Fig. \ref{fig:img-diagram}.

\subsubsection{Fingertip detection}
\begin{figure}[htbp]
  \centering
  \includegraphics[width=\columnwidth,keepaspectratio, trim=8mm 5mm 8mm 8mm, clip=true]{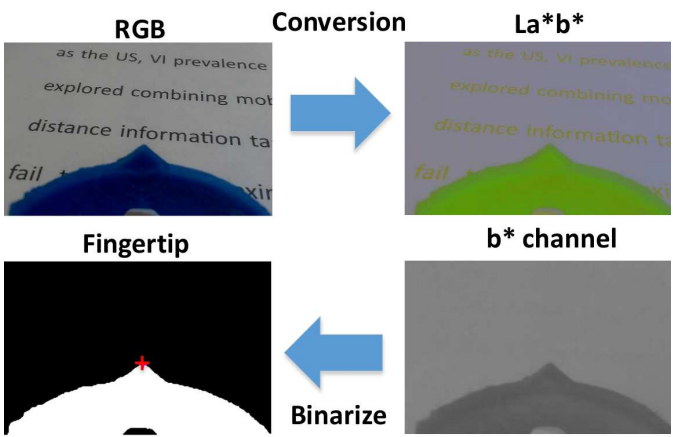}
  \caption[]{The fingertip detection process is shown here. The conversion to the L*a*b* color space helps with identifying the blue color of the fingertip device which gives the position of the fingertip.}
  \label{fig:img-fig2}
\end{figure}

\hfill\\
The device that the users wear on his finger is made of a blue resin-based material. The blue color information makes the device different from the background. We first convert the RGB input image to the L*a*b* color space. The chromaticity-layer ‘b*’ indicates where the color falls along the blue-yellow axis. Because the blue color can be visually distinguished from others in the b* channel, we can use the Otsu’s method that performs clustering-based image thresholding automatically to segment the object from the background. The fingertip then can be detected as the topmost pixel on the contour of the largest object. Fig. \ref{fig:img-fig2} shows the intermediate and final results of the fingertip detection.

\subsubsection{Text line detection}
\hfill\\
Instead of using adaptive thresholding in a sliding window for text line detection \cite{shilkrot2015fingerreader}\cite{stearns2016evaluating}, our method applies the FAST algorithm \cite{rosten2006machine} to detect corners quickly in images. This is done to speed up the processing. We also use the binary mask generated in the previous stage to remove corners that are not on the page surface. As a result, the corners in image are only from text lines. 

\begin{figure}[htbp]
  \centering
  \includegraphics[width=\columnwidth,keepaspectratio, trim=8mm 8mm 8mm 8mm, clip=true]{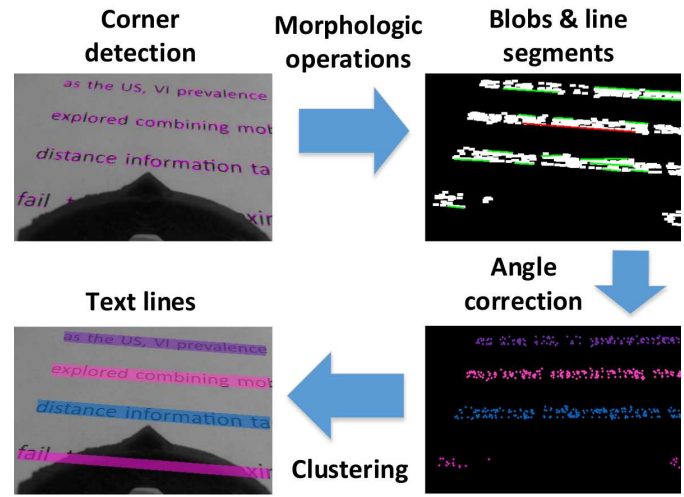}
  \caption[]{The text line detection process is shown here. The FAST algorithm was used to extract the corners. The morphologic operations aggregate the corner blobs.}
  \label{fig:img-fig3}
\end{figure}

The corners may be sparse in the image. We need to aggregate them for facilitating the consequent processes. The morphologic operations are used to process corners for generating many binary blobs in the image. Then, the FAST line detection algorithm \cite{lee2014outdoor} is applied to find the line segments in the binary image. The angle of the longest line segment is selected as the angle of whole text block. This angle is then used to rotate the corner image so that the principle angle of the corner distribution aligns with horizontal direction. Lastly, these rotated corners can be clustered to several groups according to their y-coordinate values. The group of corners are transformed inversely to original image to form the regions of text lines. Fig. \ref{fig:img-fig3} shows an illustration of the text line detection.

\subsubsection{Word recognition}
\begin{figure}[htbp]
  \centering
  \includegraphics[width=\columnwidth,keepaspectratio, trim=9mm 8mm 9mm 8mm, clip=true]{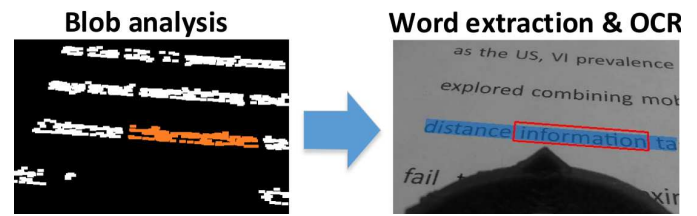}
  \caption[]{The word extraction process is shown here. The connected component analysis is used to find the largest blob close to the fingertip. A search based on the bounding boxes overlapping from other blobs finds the word.}
  \label{fig:img-fig4}
\end{figure}

\hfill\\
By acquiring the positions of the fingertip and the text lines, we can move forward to the next task which is the word recognition. The first step is to extract the focused word from a text line. 
Again, we use the blob image that was obtained in the previous stage to start this process. The Connected Component Analysis is executed for blobs in the text line above and closest to the fingertip. The largest blob of this text line is selected. Then, we search for other blobs whose bounding box are overlapping with the largest blob, and finally merge them to form the word blob. The next step is to crop the word region from the original gray image according to the bounding box of the word blob. Finally, the Tesseract OCR engine \cite{smith2007overview} is used for word recognition. Tesseract-OCR is considered as one of the most accurate open source OCR engines currently available. Tesseract handles input test image in the grayscale or binary format and follows a traditional step-by-step pipeline for the OCR. Fig. \ref{fig:img-fig4} shows an illustration of the word extraction. 

\subsubsection{Finger motion on the page surface}
\hfill\\
The motion estimation of finger on the page surface can provide information such as velocity and orientation, which are useful for the analysis of finger movement. Optical flow is 2D vector field where each vector is a displacement vector showing the movement of points between two consecutive frames. Because the camera is mounted in the finger of a user, we instead compute the optical flow of text to estimate its relative velocity to the finger/camera.   
\\
To compute the optical flow between two images, we have to solve the following constraint equation:
\begin{equation}
\label{eq:flow}
f_x u + f_y v + f_t =0
\end{equation}
where $f_x$ and $f_y$ are image gradients, $f_t$ is the gradient along time. Unknown variables $u$ and $v$ are the horizontal and vertical optical flow. The method used here for solving (\ref{eq:flow}) is based on the Lucas-Kanade method. Here we divide the original image into many $3 \times 3$ patches. For each patch, the nine points inside are assumed to have a constant velocity. So now the problem becomes solving nine equations with two unknown variables, which is over-determined. The method performs a weighted least-square fit for $u$ and $v$. The final solution is as follows:
\begin{equation}
\label{eq:Lucas}
\begin{bmatrix}
u\\
v
\end{bmatrix}=
\begin{bmatrix}
\Sigma_{i} f_{x_i}^2		& 	\Sigma_{i} f_{x_i}f_{y_i}	\\
\Sigma_{i} f_{x_i}f_{y_i}	&	\Sigma_{i} f_{y_i}^2
\end{bmatrix}^{-1}
\begin{bmatrix}
-\Sigma_{i} f_{x_i}f_{t_i} 	\\
-\Sigma_{i} f_{y_i}f_{t_i}
\end{bmatrix}
\end{equation}
The inverse matrix in (\ref{eq:Lucas}) is similar to the Harris corner detector, which denotes that corners are suitable for being tracked by optical flow. In the initial step, the most prominent corners are detected by Shi-Tomasi corner detector, and then in the second step the Lucas-Kanade optical flow method iteratively tracks these points. The two steps are executed repeatedly until frame capturing stops. 
\\
The motion orientation can be computed by using corner points and their optical flow. Suppose that $src$ is the point set detected in the initial step and $dst$ is the point set tracked by optical flow. The affine transformation from $src$ to $dst$ contains the translation that can be equivalent to the motion orientation of the point set.
\\
Assume that the $2 \times 3$ matrix $[A|b]$ is the affine transformation, where the $2 \times 2$ matrix $A$ encodes rotation, scaling, shearing, and the vector $b$ encodes translation. The problem is formulated as follows:
\begin{equation}
\label{eq:aff}
[A^* | b^*] = \argminD_{[A|b]} \Sigma_{i} \Vert dst[i] - A.src [i]^T - b \Vert ^2
\end{equation}
The minimization of (\ref{eq:aff}) then can be converted into a least squares problem. Our results show a perfect translation vector obtained from the estimation of affine transformation, which is considered as the motion orientation of finger on the page surface.

\subsubsection{Feedback for finger movement}
\begin{figure}[htbp]
  \centering
  \includegraphics[width=0.8\columnwidth,keepaspectratio, trim=8mm 8mm 8mm 8mm, clip=true]{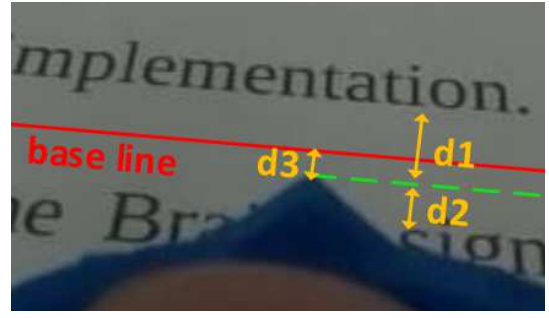}
  \caption[]{The three distances for feedback calculation: $d1$ is the distance between the baseline and the upper text line, $d2$ is the distance between the baseline and the lower text line, and $d3$ is the distance between the baseline and the fingertip.}
  \label{fig:img-fig5}
\end{figure}

\hfill\\
The motion estimation of the fingertip on the page surface was also needed to provide information such as velocity and orientation, which are useful for the analysis of the fingertip movement.
It is essential to design a mechanism that can use the visual information to form a feedback value. This feedback is used for guiding the BVI users for tracing their fingers over the text lines. We draw a baseline to equally split the area between the two text lines. Then, we calculated three distances: $d1$ is the distance between the baseline and the upper text line, $d2$ is the distance between the baseline and the lower text line, and $d3$ is the distance between the baseline and the fingertip. These are shown in Fig. \ref{fig:img-fig5}.

We can define the feedback strength, $s$, as follows:

\begin{equation}
\label{eq:feedback}
\begin{gathered}
	s=(-1)^k (\frac{d_3}{d_3 + min(d_1, d_2)}), \\[3mm]
	\hspace*{30mm}  k=
	\begin{cases}
    0,		& \text{if fingertip is above baseline}\\
    1,      & \text{otherwise}
	\end{cases}
\end{gathered}
\end{equation}

\begin{figure}[htbp]
  \centering
  \includegraphics[width=\columnwidth,keepaspectratio, trim=8mm 5mm 8mm 5mm, clip=true]{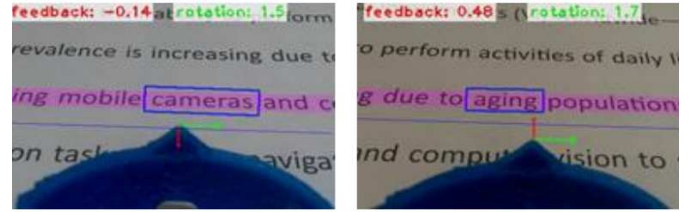}
  \caption[]{The two cases for finger moving up and down with feedback. The user will move the finger down if the sign is positive or up if the sign is negative. This feedback is sent to the E-Braille display.}
  \label{fig:img-fig6}
\end{figure}

The sign of $s$ indicates the direction of movement of the finger. The users will move their finger down if the sign is positive or up if the sign is negative. The absolute value of $s$, which is in the range of [0,1], indicates the strength of signal for E-braille display. Fig. \ref{fig:img-fig6} shows two cases where a finger is moving above and below the baseline with the feedback calculated from the positions of the fingertip and the baseline.

By using the feedback mechanism shown in (\ref{eq:feedback}), the BVI users will be able to move their finger around the baseline; they can receive the proper feedback from the E-braille display when their finger is moving away from the baseline.

Also, by calculating the distance between the fingertip and the endpoints of a text line, we can
determine if the fingertip is located near the beginning of the text line or is approaching the end of the line.

In our current application, the system can process two to three frames of video per second.

\subsubsection{Algorithm}
\hfill\\
There are multiple different algorithm that work in conjunction with each other for various parts of the system. The codes are written in Python language and are over thousands of lines. Some parts are basically an implementations of the equations presented here. There are other parts to control tens of relays, control the amperage and power of the electrical signal, translating characters to Braille and subsequently to the electrical signal, controlling the display, feedback, log recording, and etc.
\\
Here in Algorithm 1, a simplified sudo-code is presented to show the overall picture of the algorithm.
\begin{algorithm}
\label{navalgo}
 \caption{Simplified sudo-code for text extraction and feedback algorithm }
  \begin{algorithmic}[1]
 \STATE{\emph{\textbf{Algorithm:}}}\\
 \STATE{$Initialization\leftarrow$\nomenclature{$Initialization$} Initialize the camera, control board, and relays}
 \STATE{\emph{$\backslash*$ There are input signals that need to be initialize individually$*\backslash$}}

  \FOR{$Frame$ of $CameraCapure$}
  \STATE{$CameraCapure\leftarrow$\nomenclature{$CameraCapure$} The camera captures the first image}
  \STATE{$R-OCR\leftarrow$\nomenclature{$R-OCR$}{Center} R-OCR is performed on the frame}
  \STATE{$Lines\leftarrow$\nomenclature{$Lines$}{CenterPSD} Text lines are extracted}
    \FOR{$Word$ of $Lines$}
          \IF{$L_d <= L_{criteria}$}
           \STATE{\emph{$\backslash*$ $L_d$ being the distance from line $*\backslash$}}
          \STATE Process $Word$ with R-OCR.
                \FOR{$C_w$ of $Word$}
                    \STATE Process $C_w$ to match with Braille signal.
                    \STATE Open relays corresponding to $C_w$.
                        \IF{$i$ !within $i_{criteria}$}
                            \STATE{\emph{$\backslash*$ $i$ being the amperage value of signal $*\backslash$}}
                            \STATE Regulate the current to match $i_{criteria}$.
                        \ENDIF
                \ENDFOR
                \IF{$Word$ is last of $Lines$}
                 \STATE Send signal to move to a new line
                 \ENDIF
          \ENDIF
          
          \IF{$L_d > L_{criteria}$}
                \STATE Analyze the corrective action.
                \STATE Send signal to current finger position.
          \ENDIF
  
    \ENDFOR

 \ENDFOR

  \end{algorithmic}
 \label{BiSecalgorithm}
 \end{algorithm}

\section{Experimental Procedures and Setup}
\label{sec:validation}
\subsection{Participants}
In the experiments, six subjects (4 males and 2 females) participated. They were all healthy graduate students and researchers at the University of Nevada, Reno. There were no reports of any physical problems and they neither had any complications nor showed signs of any issues with their skin. 
At this stage, all participants were sighted but were blindfolded for the purpose of the experiments. 
The future experiments will include recruiting BVI participants for comparison.
Participation was with informed consent and followed protocols approved by the University of Nevada, Reno Institutional Review Board.
The goal and the procedure of the experiment were explained to the participants thoroughly.

The participants were seated on a chair that was $40cm$ from the ground. The table in front of them was $68cm$ from the ground. The distance between the person and the bottom of the paper (US Letter paper) was $20cm$ and the distance from the person to the system was $30cm$. These distances were considered to make sure the participants' arms can reach the table and can move freely. It was later seen that the movements from the arm, instead of the wrist will result in significantly less corrections for up and down commands.

\subsection{Experimental Procedure}
The research was broken down into two parts. These two parts were designed for comparing the effectiveness of the feedback given by the system to track the line versus tracking the line without any feedback.

\begin{figure}[htbp]
  \centering
  \includegraphics[width=0.9\columnwidth,keepaspectratio, trim=0mm 0mm 0mm 0mm, clip=true]{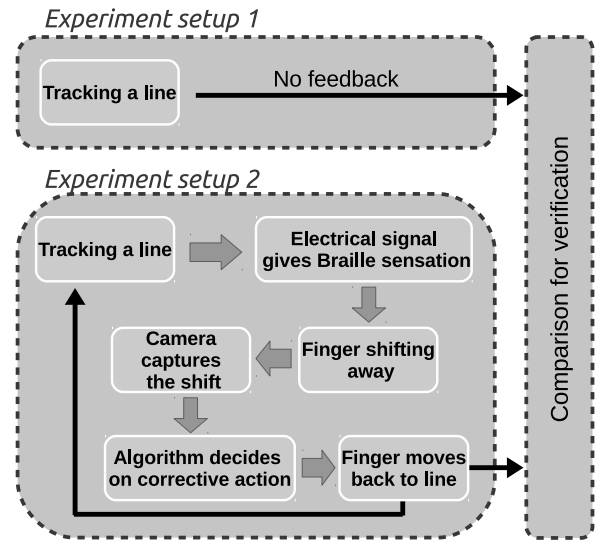}
  \caption[]{The validation was conducted as two steps. One to examine the natural movement of the fingertip and another one to examine the effectiveness of the feedback method.}
  \label{fig:paper_track_control}
\end{figure}

In the first part, a printed text was presented to the subjects and they were asked to place their fingertip at the beginning of a line. Then they were required to close their eyes and track the line without any feedbacks. A camera mounted above the paper captured the movement of the subject's fingertip. 

In the second part, subjects were introduced to the electro-tactile system. First, the detection threshold (DT) voltage was examined and found for each subject. They were allowed a short time between $5-20$ seconds to examine the feeling of the electrical signal. 
Then, the up and down commands were introduced to them as an alternating signal with half a second delays. A display also showed this transition between the up and down commands. The users were given enough time until they stated that they could easily recognize each signal. 
The next step was a test to find out whether they could positively identify each signal. In this test, a random command (either up or down) was given to the fingertip and the user was asked to identify the command. This was repeated $10$ times. If the user could not identify more than one of the commands, the procedure would be restarted again from the beginning. This happened only once for our $6$ subjects and most people were able to pass these two steps in less than $5$ minutes.

Finally, when users were comfortable in identifying the up and down commands, they were asked to place their fingertips at the beginning of a line and close their eyes and start tracking the line. As their finger started to drift from the line (either going up or down) the system gave them the appropriate command (either up or down) to correct their fingertip position. This was continued until reaching the end of the line and repeated $3$ times for each user. The whole session was captured in video via an over-the-top camera. The videos were later analyzed to determine the tracking path that each user's fingertip traveled on the page.

\begin{figure}[htbp]
  \centering
  \includegraphics[width=\columnwidth,keepaspectratio, trim=5mm 5mm 6mm 5mm, clip=true]{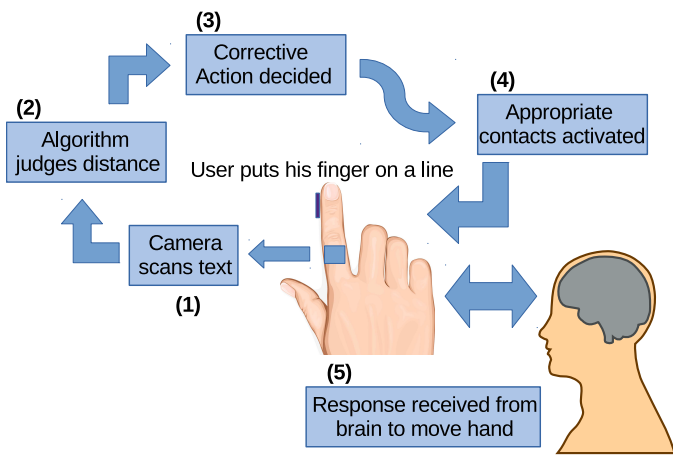}
  \caption[]{First, user puts his/her fingertip on a text line. The camera scans the text and an algorithm decides if the finger needs to be moved up or down. The appropriate contacts on the fingertip are then activated sending the signal to the nerves. The users then follow the command to adjust the fingertip position.}
  \label{fig:Block_diagram}
\end{figure}

\begin{figure}[htbp]
  \centering
  \includegraphics[width=0.5\columnwidth,keepaspectratio, trim=0mm 0mm 0mm 0mm, clip=true]{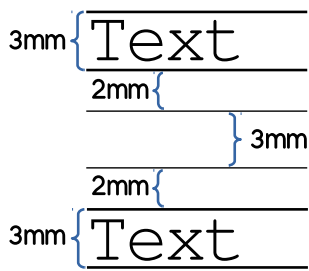}
  \caption[]{Each line of the text had a height of $3mm$ with a distance of $10mm$ from the lines above or below (center-line to center-line). The system was able to maintain users' fingertip within a $2mm$ distance above and below the line.}
  \label{fig:text}
\end{figure}

The block diagram of the procedure is shown in Fig. \ref{fig:Block_diagram}. The text on the page was set up as Fig. \ref{fig:text} to have double-spaced lines with font size of $14$.

\section{Results and Discussions}
\label{sec:discussion}

\subsection{Comparison of the Trackings With or Without Feedback}

\begin{figure}[htbp]
  \centering
  \includegraphics[width=1\columnwidth,keepaspectratio, trim=5mm 0mm 13mm 5mm, clip=true]{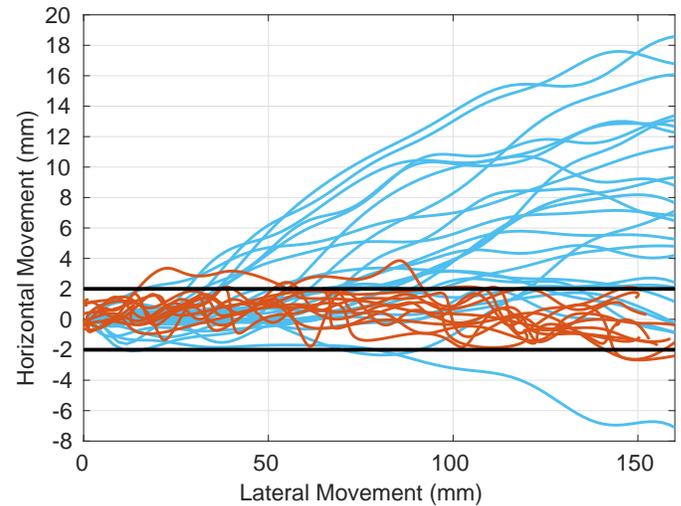}
  \caption[]{The blue lines are the natural fingertip movements in the absence of any feedback. The brown lines are the finger movements that received the up and down feedback. It is clear that the system was able to control the drifting of the fingertip and maintain its movement within $2mm$ distance of the line.}
  \label{fig:comparison}
\end{figure}

\begin{figure}[htbp]
  \centering
  \includegraphics[width=0.8\columnwidth,keepaspectratio, trim=7mm 0mm 12mm 5mm, clip=true]{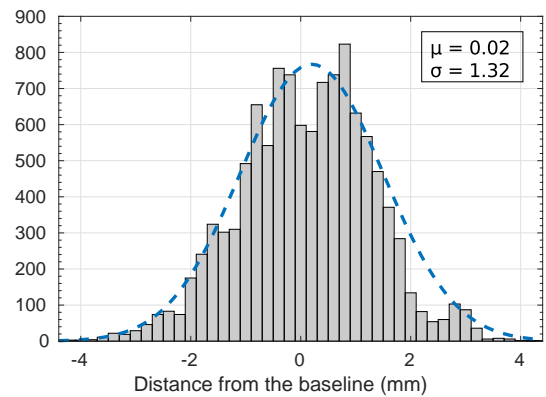}
  \caption[]{The histogram of fingertip locations when the users received control feedbacks for $11440$ data points from $25$ tests. The mean is $0.02$ and the standard deviation is $1.32$.}
  \label{fig:histo}
\end{figure}

As it was explained in the procedure section, a comparison was done between the movements of the fingertip over a text line while receiving the up and down feedback versus without any feedback presented to the user.
Fig. \ref{fig:comparison} shows this comparison. These were done while the users had their eyes closed. Fig. \ref{fig:histo} shows the histogram of fingertip location with feedback.

\begin{figure}[ht]
  \centering
  \includegraphics[width=0.90\columnwidth,keepaspectratio, trim=4mm 0mm 10mm 9mm, clip=true]{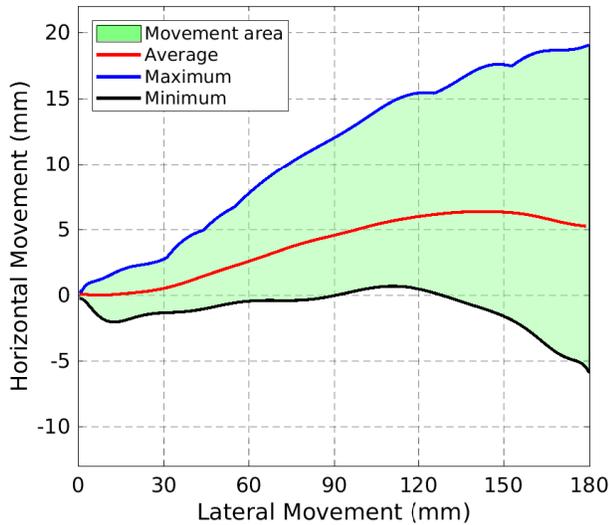}
  \caption[]{Multiple repetitions of the users tracking the line without any feedback and with their eyes closed is summarized here. The maximum and minimum drift are shown. The average of the tracking path for all users/repetitions is also shown. It is clear that even in the average path, a drift of more than $6mm$ from the line can happen.}
  \label{fig:Avg_Min_Max}
\end{figure}

As it can be seen in Fig. \ref{fig:comparison}, when users are not getting any feedback for tracking the lines, there will be a significant drift in the tracing. In some cases, it almost got to $20mm$ away from the text by the end of a typical line. This drift was different for each user. Fig. \ref{fig:Avg_Min_Max} shows how all users performed in tracking the lines without any feedback. An average of all tracking paths performed by the users shows a drift of more than $6mm$ from the line. It is important to note that the range (shown by the maximum and minimum distance) should be given more value than the average since there are tracking paths above and below the line that can cancel out each other in the average. The maximum and minimum lines show the absolute range for all of the tracking paths and not only one specific path.

\begin{figure}[ht]
  \centering
  \includegraphics[width=0.90\columnwidth,keepaspectratio, trim=7mm -5mm 10mm 5mm, clip=true]{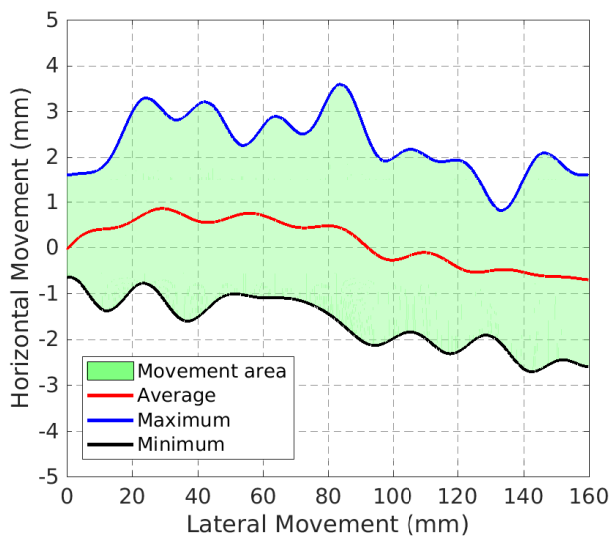}
  \caption[]{A summary of how all users performed in multiple repetitions for each user is shown here. The range of movement shown by the maximum and minimum of all the tracking shows that the system was able to maintain the user's fingertip within a $2mm$ distance of the line. The average is within less than $1mm$ distance of the line.}
  \label{fig:min_max_system}
\end{figure}

As a comparison, the same figure can be shown for when users received feedback to correct their fingertip position. Fig. \ref{fig:min_max_system} shows how users performed while receiving the up and down feedback.

As before, the maximum and minimum lines show the absolute range for all of the trackings and not only one specific path. In other words, a user may have drifted significantly up and away from the line in the beginning but did not drift at all until the end of the line, on the other hand, another user might had only a up drift at the end of the tracking. The maximum line is the combined maximum distance tracking of the users.

One important observation in Fig. \ref{fig:min_max_system} is that the lines are less smooth than the previous case. This is only because of the effect of the feedback on the movements of the fingertip. Each time the user is receiving a feedback command from the system, he or she needs to correct the fingertip position by moving to the exact opposite direction. This has resulted in continuous up and down movements along the text line.

Maintaining the fingertip location within $2mm$ distance of the line provides the camera with the best field view of the text. Since the camera is tilted and placed at an angle on the finger, the field view is significantly affected by the slightest movements of the finger. Here, keeping the fingertip within a very short distance of the line prevents such a problem and the camera can always capture the text at this distance.

\begin{figure*}[htbp]
  \centering
  \includegraphics[width=0.95\textwidth,keepaspectratio, trim=40mm 0mm 40mm 0mm, clip=true]{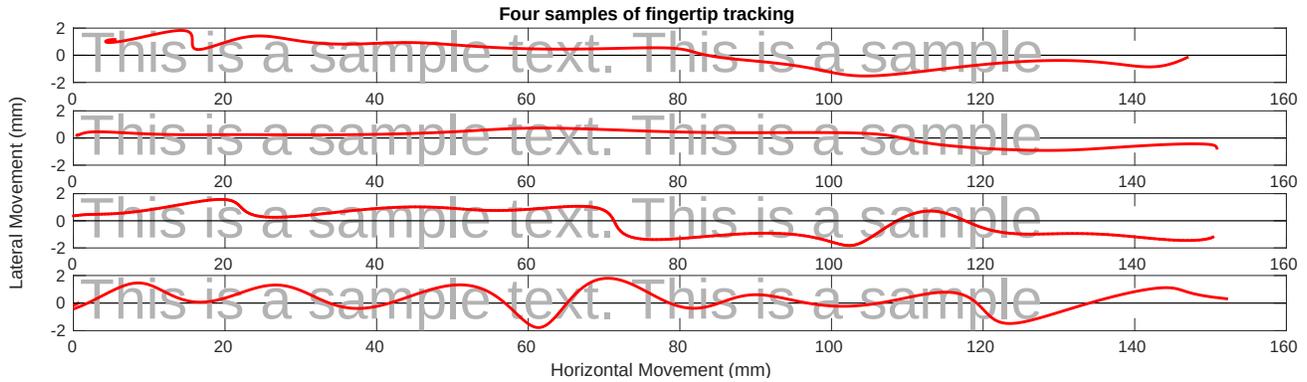}
  \caption[]{Four samples of the fingertip tracking over the text line are shown here. 
  The tracking paths were rarely as smooth as the second sample here. Most users needed several directional feedback along the line to correct their fingertip position (i.e. sample \#4).
  In the first sample, the user has slightly moved back along the line thinking that he was moving down.}
  \label{fig:four_samples}
\end{figure*}

Fours samples of the tracking of the line while receiving the feedback is shown in Fig. \ref{fig:four_samples}. Notice that at some points, the user may have moved slightly back while tracking the line.
In our analysis, it became clear very early on that the tracking paths for the experiments that the users received directional feedback are not as smooth as when they did not receive any feedback. This is because each time the user is receiving a feedback command from the system, he or she needs to correct the fingertip position resulting in constant up and down movements.

\subsection{Tracking Speed}

\begin{figure}[htbp]
  \centering
  \includegraphics[width=0.90\columnwidth,keepaspectratio, trim=4mm 0mm 14mm 1mm, clip=true]{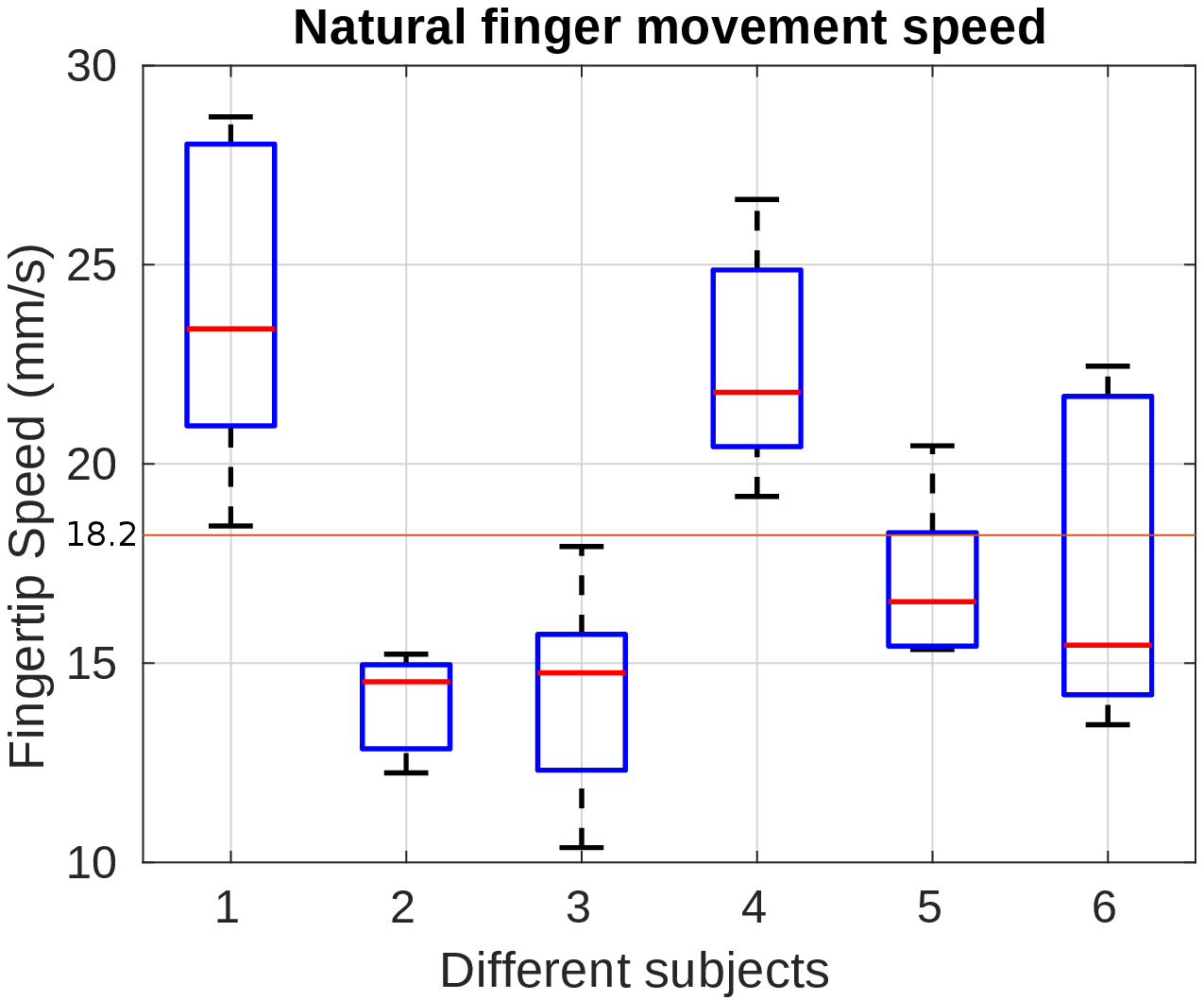}
  \caption[]{The fingertip tracking speeds for multiple repetitions of the experiment by the users without receiving any directional feedback are shown here. The average speed is $18.21mm/s$.}
  \label{fig:Natual_speed}
\end{figure}

One important indication of the performance of the system is the speed by which the users can scan a text line. 
The analysis of the tracking paths without any feedback shows an average speed of $18.21mm/s$ for scanning a text line. Fig. \ref{fig:Natual_speed} shows the speeds of the tracking for the users during multiple repetitions. This is consistent with the previous research in this area.

\begin{figure}[htbp]
  \centering
  \includegraphics[width=0.90\columnwidth,keepaspectratio, trim=4mm 0mm 14mm 1mm, clip=true]{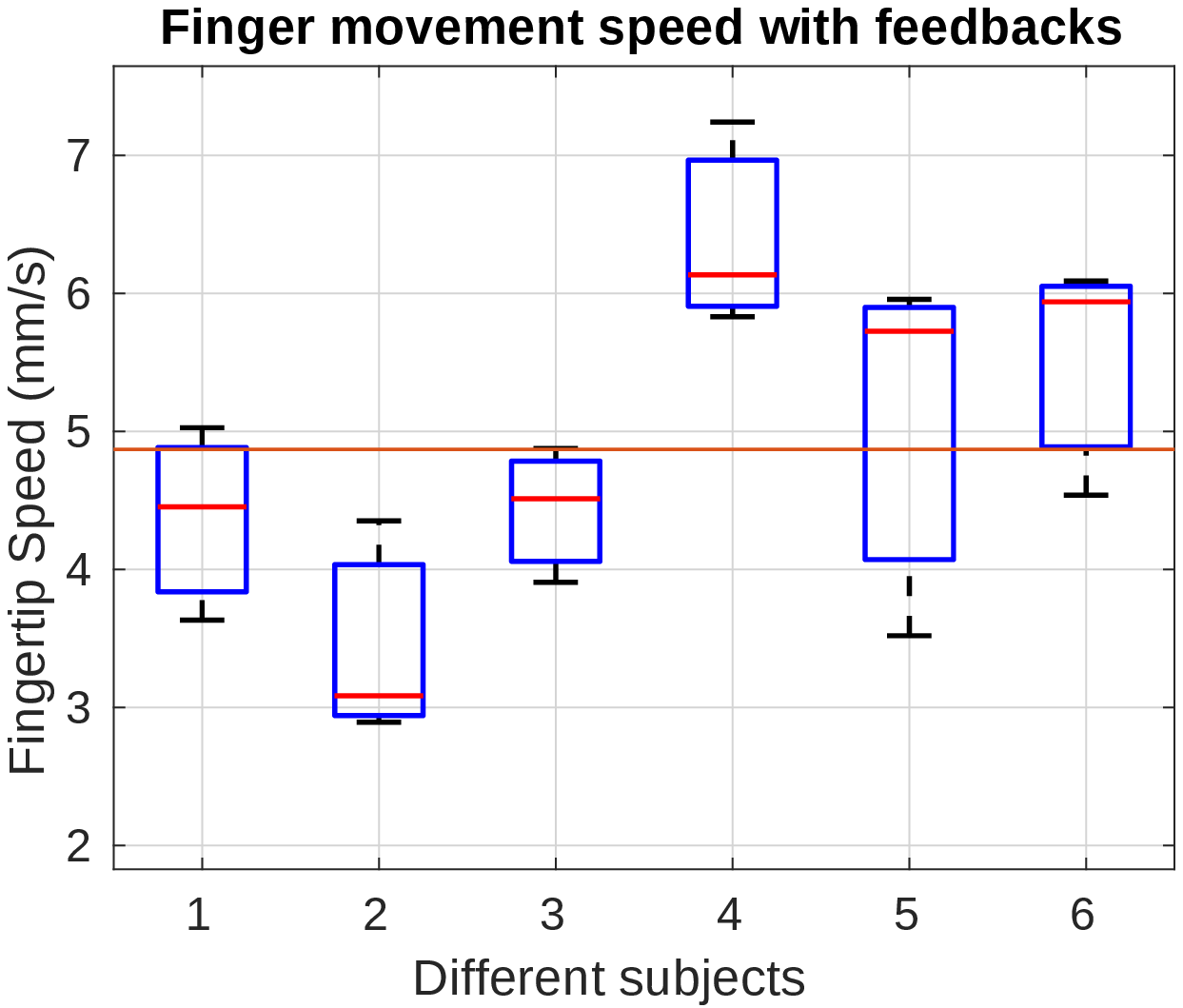}
  \caption[]{The fingertip tracking speeds for multiple repetitions of the experiment by the users while receiving the directional feedback commands are shown here. The average speed is $4.87mm/s$.}
  \label{fig:system_speed}
\end{figure}

The same analysis can be done for the experiments where the users received directional feedback commands. Fig. \ref{fig:system_speed} shows the speeds for multiple repetitions of the experiment for the users.

As it is shown, the tracking speed is significantly lower when the users receive the directional feedback commands from the system. One reason for this behavior, as was observed during conducting the experiment, is that the users tend to stop their fingertip movement as soon as they receive a feedback command. They spend a short amount of time to analyze the sensation to understand whether they are receiving an up command or a down command. For some users, this delay was around one second but at some points it could be as long as 2-3 seconds. It can be rationalized that the users are not familiar with the sensation and need to spend a short amount of time to analyze and properly identify the sensation. It can be hypothesized that this feeling would fade away as the users become more accustomed to the system.

\begin{figure}[htbp]
  \centering
  \includegraphics[width=0.95\columnwidth,keepaspectratio, trim=12mm 0mm 8mm 0mm, clip=true]{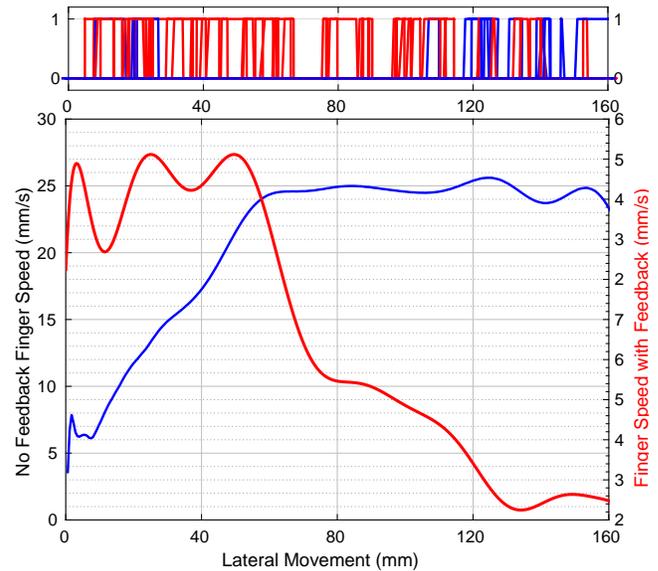}
  \caption[]{The blue line is the average speed of the natural finger movements in the absence of any feedback. The red line is the average speed of the finger movements that received the up and down feedback. Notice that the scale of the Y-axis for the blue and red plots are different. 
  
  The upper plot shows the up (blue) and down (red) commands that were sent to all users in all repetitions. Corresponding with the speed plot below, it shows that each time, a bursts of commands is followed by a drop in the fingertip speed.}
  \label{fig:Speed_old_new}
\end{figure}

The last analysis in this section is to understand whether the position of the fingertip in relation to being in the beginning of the line or at the end of the line affects the tracking speeds. In other words, how long the users spend on each part of the line. Fig. \ref{fig:Speed_old_new} presents a comparison between the average tracking speeds for all users and all repetitions for receiving the feedback commands versus not receiving any feedback. Notice that the scale of the Y-axis is different for each plot as the speeds are different. A plot of the number and location of the up/down commands is added to the Fig. \ref{fig:Speed_old_new} to show the correlation between the number of commands and the speed of the fingertip. It can be seen that in each instance, a drop in the speed of the fingertip is followed after a bursts of up/down commands.

Fig. \ref{fig:Speed_old_new} shows that in the absence of any feedback, users tend to start the tracking slowly but their tracking speed increases as they get closer to the end of the line. On the other hand, when users are receiving the feedback commands, they tend to perform the exact opposite. They start very fast and then slow down as they progress. This can be explained by the fact that in the beginning, the users put their fingertip on the line and have very small drift as they move their finger; but as soon as they drift outside of the acceptable zone and receive the feedback commands, they slow down more and more to correct their fingertip position.

\subsection{Reaction Time}

\begin{figure}[htbp]
  \centering
  \includegraphics[width=0.89\columnwidth,keepaspectratio, trim=9mm 10mm 13mm 10mm, clip=true]{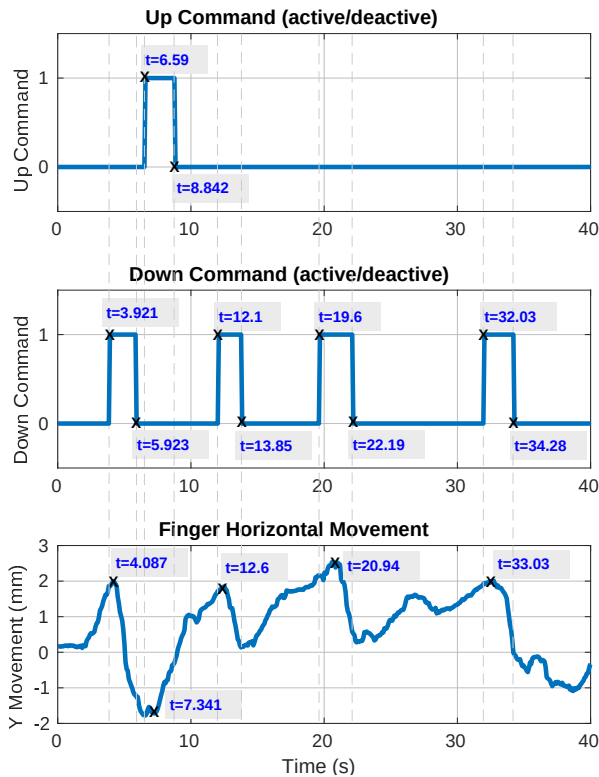}
  \caption[]{A sample of the commands and the tracking is shown for one user. The up and down commands are shown as binary signals (whether on or off). The tracking is shown only along the Y-axis (horizontal movement) to show the matching with the exact time of the commands.}
  \label{fig:up_down_commands_results}
\end{figure}

The users change their fingertip position based on the commands that they receive from the system. It was said before that the users need to spend some amount of time to analyze the sensation and then react based on the command they received. This reaction time can be seen and measured in the performance plots of each tracking. Fig. \ref{fig:up_down_commands_results} shows these commands and the user's reaction times for one sample.

Fig. \ref{fig:up_down_commands_results} can be first analyzed from the perspective of how long it takes for the user to understand and comply with the up and down commands. In this case, there is only one up command issued at $t=6.59s$. The corresponding response occurred at $t=7.341s$. This shows the user spent $751ms$ to understand and comply with the command. There are four down commands as the user is showing a tendency to drift upward. The first down command happens at $t=3.921s$ and the respond happens at $t=4.087s$. This shows a very fast response of $166ms$. The following commands and responses have the delays of: $500ms$, $1.340s$, and $1s$. It can be seen that even in this example, the response time increases as the user moves more toward the end of the line. This confirms Fig. \ref{fig:Speed_old_new} conclusions.

Another perspective to analyze this figure is to measure how long it takes the user to move his fingertip back to the designated zone (where the camera can read the text and he no longer receives any up or down commands). In this figure, the up command started at $t=6.59s$; the user begins moving his fingertip up until at $t=8.842s$ the up command stops indicating that the user is now inside the designated zone and can continue the tracking of the line. This means $2.252s$ of upward movement that subsequently results in slower tracking of the line. The duration of complying with the down commands were: $2.002s$, $1.75s$, $2.59s$, and $2.25s$. This shows an average of almost two seconds to comply with each command.

\section{Conclusion}
\label{sec:conclusion}
Although there are existing technologies that allow the blind and visually impaired, BVI, to download and translate various books and literature to Braille or as audio. However there are still significant printed books and articles that do not have available audio or Braille translations. This situation limits the amount of resources that are available to the BVI and thus, their independence. This paper focused on the research that will not only solve this problem, but also can be further extended to aid the BVI user. 
\\
In this work, based on a custom-built finger-werable electro-tactile based refreshable Braille system, a new opto-electrotactile feedback method was proposed and developed to allow tracking and following the printed text lines and then reading them by user's fingertip. The method combining electrotactile stimulation with the developed rapid Optical Character Recognition (R-OCR) approach facilitates the new text line tracking control strategy that enables text line following by user fingertip during reading. Extensive experiments were designed and conducted to test the ability of blindfolded participants to read through and follow the text line based on the opto-electrotactile-feedback method. The experiments show that as the result of the opto-electrotactile-feedback, the users were able to maintain their fingertip within a $2mm$ distance of the text while scanning a text line. The tracking speeds and the response times of the users in multiple repetitions were also examined. It was shown that the users tend to scan slower when they receive the directional commands. Also, the users needed almost one second to analyze and respond to the commands. It can be found that users with more experience with the system will take significantly less time to react to the commands. Our work is a significant step to aid the BVI users with a portable means to translate and follow to read any printed text to Braille, whether in the digital realm or physically, on any surface.
To the best of our knowledge, this work is the first-ever synergetic effort to tackle the challenges from three angles (i.e., perception, computer vision, and human-device interaction) that will produce a holistic and intelligent solution for a complete reading aid for BVI.

\bibliographystyle{IEEEtran}
\bibliography{PaperTrack}

\end{document}